\documentclass[twocolumn,aps,prl,showpacs,superscriptaddress]{revtex4}
\usepackage{graphicx}
\usepackage{amsmath,amssymb}
\usepackage{color}

\newcommand{\beq}{\begin{equation}}
\newcommand{\eeq}{\end{equation}}
\newcommand{\beqa}{\begin{eqnarray}}
\newcommand{\eeqa}{\end{eqnarray}}
\newcommand{\ket} [1] {\vert #1 \rangle}
\newcommand{\bra} [1] {\langle #1 \vert}

\def\ket#1{|\,#1\,\rangle}
\def\bra#1{\langle\, #1\,|}

\def\opone{\leavevmode\hbox{\small1\kern-3.8pt\normalsize1}}

\begin{document}


\title{Experimental multi-location remote state preparation}


\author{}
 \affiliation{Department of Physics, Stockholm University, S-10691,
 Stockholm, Sweden}

\author{Magnus R\aa dmark}
 \affiliation{Department of Physics, Stockholm University, S-10691,
 Stockholm, Sweden}

\author{Marcin Wie\'sniak}
 \affiliation{Instytut Fizyki Teoretycznej i Astrofizyki, Uniwersytet
Gda\'{n}ski, PL-80-952 Gda\'{n}sk, Poland}

\author{Marek \.{Zu}kowski}
 \affiliation{Instytut Fizyki Teoretycznej i Astrofizyki, Uniwersytet
Gda\'{n}ski, PL-80-952 Gda\'{n}sk, Poland}

\author{Mohamed Bourennane}
 \affiliation{Department of Physics, Stockholm University, S-10691,
 Stockholm, Sweden}


\date{\today}


\begin{abstract}
Transmission of quantum states is a central task in quantum information science.  Remote state preparation (RSP) has the same goal as teleportation, i.e. transferring quantum information without sending physically the information carrier, but in RSP  the sender knows the state which is to be transmitted. We present  experimental demonstrations of RSP for two and three locations.
In our experimental scheme Alice (the preparer) and her three partners share  four and six photon polarization entangled singlets. This allows us to perform RSP of two or three copies of a single qubit states, a two qubit Bell state, and a three qubit $W$, or $\overline{W}$ state. A possibility to prepare a two-qubit non-maximally entangled and GHZ states is also discussed. The ability to remotely prepare an entangled states by local projections at Alice 	is a distinguishing feature of our scheme.
\end{abstract}

\maketitle
Theoretical studies in quantum information predict existence of various types of entangled states, which could be useful in many communication situations, and information processing,
 for example, quantum key distribution \cite{Ekert91}, teleportation \cite{bennett93}, etc.
 Correlations between entangled systems are so strong that they cannot be modeled by any classical means \cite{Bell66}. In theory we can study entangled states of very many qubits, and complicated quantum protocols. But experimental practice shows that protocols involving many qubits are very difficult to demonstrate in the laboratory. In order to see to what extent theoretical quantum information science talks about experimentally controllable phenomena, one has to keep on testing the limits of the range of feasibility of such schemes, and keep extending such limits.
With this in mind, we present realizations of several remote state preparation (RSP) protocols, \cite{P00,BBSSTW01, LS03, BS03}, using tools of advanced multiphoton quantum interferometry \cite{PAN}.

The aim of teleportation and RSP is to take the advantage of entanglement  to prepare a desired state at a distant location. In teleportation protocol, Alice's task is to prepare an unknown, given to her, quantum state  at Bob's location.
In the case of RSP Alice knows which state she wants to prepare at Bob's location. The basic, most elementary scheme runs as follows.
Alice and Bob share a maximally entangled state of two qubits, say singlet. Alice performs a projective measurement in a basis, which contains the state which is to be remotedly  prepared. If her measurement locally projects onto the state orthogonal to the one she wants to prepare, Bob's sub-system collapses into the required state.  She sends a single bit \cite{P00} announcing whether or not her projection measurement was successful.  Such an experiment was realized with polarization qubits \cite{PNGWK05} and with an photon-atom system \cite{RBVWW07}. Note, that such a protocol must be probabilistic. Alice has a probability of $\frac{1}2$ of projecting onto the required state. In the case of failure Bob obtains a state orthogonal to the intended one. Because of the  impossibility of a universal NOT gate, such a state cannot be corrected without the knowledge of the basis to which belongs. Nevertheless, if Alice is choice restricted to e.g. states from the equator of the Bloch sphere, the protocol becomes deterministic. Simply, given the bit from Alice, Bob may perform the $\sigma_z$ operation, which acts as the NOT gate on the equatorial plane.

In this letter present a more general scheme, allowing Alice to remotely prepare a large class of symmetric states, including entangled ones. For this purpose we will utilize rotationally invariant multi-qubit singlet states.

We begin with a brief description of the experimental set-up which allows to prepare such generalized singlet states, using methods of multiphoton interferometry.  The setup consists of a non-linear crystal allows an efficient down-conversion process (non-collinear type-II PDC).  Photons from a pulsed laser pumping field can spontaneously, with a  low probability, fission  into a pair of photons with orthogonal polarizations, in two conjugate propagation modes. If pumping is strong enough one can observe multi-fold emissions of such kind form a single pulse.
The state
can be expressed as
\begin{eqnarray}
\ket{PDC} &=& \frac{1}{\cosh^2 K}\sum_{p=0}^\infty \tanh^p K \sum_{m=0}^p e^{im\phi} \\ \nonumber
&&\ket{mH_a, (p-m)V_a,(p - m)H_b,mV_b}
\label{pdc}
\end{eqnarray}
where,  $\ket{nX_c}$ denotes a Fock state with $n$  photons, of polarization $X=H,V$ in mode $c=a,b$. The parameter $K$ is a function of the non-linearity and length of the crystal,
pump power and filtering bandwidth, and $\phi$ is the possible phase difference between horizontal and vertical polarization due to birefringence in the crystal \cite{LWZBW10}. The $n$-th order PDC emission  corresponds to terms with  $p=n$.
The trick is to  place  $n-1$ consecutive beam splitters in each of the two emission spacial modes, and observe $2n$-fold coincidences, \cite{WZ}.
Correlations characteristic for four and six (polarization) qubit states ($|\Psi_k^{-}\rangle,k=,2,3$) can be observed in this way . When represented in the $H/V$ polarization  basis, these states are
\begin{eqnarray}
  &\ket{\Psi_{4}^{-}} = \frac{1}{\sqrt{3}}\Big((\ket{HHVV}+\ket{VVHH})&\nonumber\\
   &-\frac{1}{2}(\ket{HVHV}+\ket{HVVH}&\nonumber\\
&
+\ket{VHHV}+\ket{VHVH})\Big),&\\ \nonumber
   \label{Psi4}
\end{eqnarray}
and
\begin{eqnarray}
   &\ket{\Psi_{6}^{-}} = \frac{1}{2}(\ket{HHHVVV}-\ket{VVVHHH})&\nonumber\\
   &+\frac{1}{6}(-(\ket{VHH}+\ket{HVH}+\ket{HHV})&\nonumber\\
   &\otimes(\ket{VVH}+\ket{VHV}+\ket{HVV})&\nonumber\\
   &+(\ket{VVH}+\ket{VHV}+\ket{HVV})&\nonumber\\
   &\otimes(\ket{VHH}+\ket{HVH}+\ket{HHV})).&
   \label{Psi6}
\end{eqnarray}
The four-qubit state was  reported in Ref. \cite{GBEKW03}, while R\aa dmark et al. \cite{RZM09} observed the   six-qubit one.
The states are generalizations of singlets, that is they have the same form irrespective which pair of orthogonal (in general elliptic) polarizations is used to express the polarization of {\em each and every} qubit. This implies rotational symmetry: if each qubit is rotated by the same unitary transformation $U$, such that $\det U=1$, the states does not change,
$U^{\otimes k}\ket{\Psi_k^-}=\ket{\Psi_k^-}$ just like the two qubit singlet $\ket{\Psi_2^-}$.
This property can be used to circumvent some forms decoherence, \cite{Z91}. If the interaction with the environment is symmetric under an exchange of systems, one can process information within a so-called decoherence subspace,  \cite{ZR97a, LCW98}. For a decoherence process, in  a form a random rotation, acting of all qubits in the same way, such a space is spanned by singlet states. For four qubits such a decoherence-free subspace is spanned by two orthogonal four-qubit states invariant under such transformations. One of them describes the product of two two-qubit singlets $\ket{\psi_2^-}\otimes \ket{\psi_2^-}$ and the other one is $\ket{\psi_4^-}$. A decoherence-free  operation in this subspace has been demonstrated experimentally  in Ref.  \cite{BEGKCW04}.

All reduced density operators of the subsystems in $\ket{\Psi_k^-}$ are also rotationally invariant:
\begin{eqnarray}
\rho&=&\text{Tr}_S\ket{\Psi_k^-}\bra{\Psi_k^-}\nonumber\\
&=&\text{Tr}_SU^{\otimes k}\ket{\Psi_k^-}\bra{\Psi_k^-}U^{\dagger\otimes k}\nonumber\\
&=&\text{Tr}_SU^{\otimes k\setminus S}\ket{\Psi_k^-}\bra{\Psi_k^-}U^{\dagger\otimes k\setminus S}\nonumber\\
&=&U^{\otimes k\setminus S}\rho U^{\dagger\otimes k\setminus S},
\end{eqnarray}
where $S$ stands for the traced out part of the system and $U^{k\setminus S}$ is a tensor product unitary acting only on all non traced out qubits. In the third line we used the fact that trace operation  is basis-independent.

Using such states Alice can, by projecting her half of the qubits (which originate from one of the propagation modes of the PDC radiation), efficiently change the state of remote qubits (from the other propagation mode, sent to her Partners). Here we consider remote state preparation with $|\Psi_2^-\rangle$, $|\Psi_4^-\rangle$, and $|\Psi_6^-\rangle$.

To make our discussion more transparent, we can put $\ket{\Psi_2^-}$, $\ket{\Psi_4^-}$ and $\ket{\Psi_6^-}$  as
\begin{eqnarray}
\ket{\Psi_2^-}&=&\frac{1}{\sqrt{2}}(\ket{\psi}\ket{\overline{\psi}} - \ket{\overline{\psi}}\ket{\psi}),\nonumber\\
\ket{\Psi_4^-}&=&\frac{1}{\sqrt{3}}(\ket{\psi\psi}\ket{\overline{\psi}\overline{\psi}}+\ket{\overline{\psi}\overline{\psi}}\ket{\psi\psi})\nonumber\\
&+&\frac{1}{\sqrt{6}}(\ket{\psi\overline{\psi}}+\ket{\overline{\psi}\psi})\ket{\Psi_2^+}\\
\ket{\Psi_6^-}&=&\frac{1}{2}(\ket{\psi\psi\psi}\ket{\overline{\psi}\overline{\psi}\overline{\psi}}-\ket{\overline{\psi}\overline{\psi}\overline{\psi}}\ket{\psi\psi\psi}),\nonumber\\
&+&\frac{1}{2\sqrt{3}}(\ket{\psi\overline{\psi}\overline{\psi}}+\ket{\overline{\psi} \psi\overline{\psi}}+\ket{\overline{\psi}\overline{\psi}\psi})\ket{W_3}\nonumber\\
&-&\frac{1}{2\sqrt{3}}(\ket{\overline{\psi}\psi\psi}+\ket{\psi\overline{\psi}\psi}+\ket{\psi\overline{\psi}\overline{\psi}})\ket{\overline W_3},\\ \nonumber
\end{eqnarray}
where $\ket{W_3}=\frac{1}{\sqrt{3}}(\ket{\psi\psi\overline{\psi}}+\ket{\psi\overline{\psi}\psi}+\ket{\overline{\psi}\psi\psi})$, while
$\ket{\overline W_3}=\frac{1}{\sqrt{3}}(\ket{\overline{\psi}\overline{\psi}\psi}+\ket{\overline{\psi}\psi\overline{\psi}}+\ket{\psi\overline{\psi}\overline{\psi}})$, and $\ket{\Psi_2^+}=(\ket{\psi\psi}+\ket{\overline{\psi}\overline{\psi}})$. Due to the rotational invariance $\psi$ and $\overline{\psi}$ may denote any pair of orthogonal polarizations.

In order to have one universal setup for remote state preparation employing the above states, the pumping parameter must be such that the emission of a single pair is approximately by an order of magnitude more probable then for emission of two pairs. This automatically guarantees that probability of three pair emission is lower by yet another order of magnitude.
Such conditions  allow high interferometric contrast (visibility) in  two, four, and six fold coincidence detections (the interference occurs while one changes the polarization settings at final analyzers at each of the exit arms of the beam-splitter system), see \cite{LWZBW10}.  Note,  that lower pump rates, could make the contrast higher, but the count rates of six-fold coincidence detections would become prohibitively low. Thus a proper tuning of the pump strength must be made.

The interesting feature of the four-beam-splitter setup of Fig. 1  is that whenever Alice registers just a single count in one of her three detection stations, under the pumping conditions described above, the most probable events on the other side of the interferometer (shared by Bob, Charlie and David, each controlling one of exits) are no counts at all (due to an inefficiency of the detectors), or just a single count at one of the three exit arms (two photon event probability is lower by an order of magnitude times efficiency of the detectors, which is close to two orders of magnitude).
Thus, Alice is able
to remotely prepare any pure single-qubit state for her partners, but she does not have control who actually receives it. To prepare  $\ket{\psi}$, Alice sets measurement stations, all three,  to the $\ket{\psi}/\ket{\bar\psi}$-basis.
Every time Alice gets the result $\ket{\bar\psi}$ in
 one of her stations, while other two do not register anything, one her partners will have the state $\ket{\psi}$. However, if she registers in her all three measuring stations photons of the same polarization, she is (almost) sure that if all her partners register photons, then these will of the same polarization, orthogonal to the one she measured at each of the stations, see the fist term  of $\ket{\Psi_6^-}$ . If she gets just two counts, at different stations, with highest probability a two pair emission occurred, thus, we have the case of $\ket{\Psi_4^-}$, and she can be (almost) sure that a pair of her partners, if they register single photons,  have qubits of polarization $\ket{\overline\psi}$, but its is beyond her control who gets them.

Alice can also conditionally
 prepare a three-qubit entangled state  $\ket{W_3}$ or  $\ket{\overline{W}_3}$ for her partners to share.  Alice  measures at all her stations  in the basis $\{\ket{\psi}, \ket{\overline{{\psi}}}\}$. If she gets a count at each of her stations, consistent with three-qubit states  $\ket{\psi\overline{\psi}\overline{\psi}}$, $\ket{\overline{\psi}\psi\overline{\psi}}$ or $\ket{\overline{\psi}\overline{\psi}\psi}$, the remote parties will be sharing  $W$ state, provided each of them received just one photon.  If she registers  $\ket{\psi\psi\overline{\psi}}$, $\ket{\psi\overline{\psi}\psi}$ or $\ket{\overline{\psi}\psi\psi}$, the $\overline W$ state is remotely prepared (under the same proviso).

Similarly, if we have a two pair emission leading to  the $\ket{\Psi_{4}^{-}}$ 
state, Alice can prepare the Bell state $\ket{\Psi^+}$, shared by a pair of her Partners. It is so provided  Alice measures $\ket{\psi\overline{\psi}}$ or $\ket{\overline{\psi}\psi}$ at a pair of her stations (and no counts at the third station), and two partners receive (register) photons.

It is important to notice that operating on $\ket{\Psi_{6}^{-}}$ Alice can prepare  genuinely three-partite entangled pure states $W$ and $\overline W$, by just using projections onto factorizable states. Interestingly, to  prepare a Greenberger-Horne-Zeilinger state (GHZ), she needs to register one of her qubits in state $\ket{\psi}$,  second  in state $\cos{\theta}\ket{\psi}+\sin{\theta}\ket{\overline{\psi}}$, and the third one in  $\cos{\theta}\ket{\psi}-\sin{\theta}\ket{\overline{\psi}}$, where $\theta=\pm\frac{\pi}{3}$. A back of an envelope calculation shows that in such a case, state $\ket{\Psi_6^-}$ collapses in such a way that  Bob, Charlie, and David share $\frac{1}{2}(\ket{\overline{\psi}\overline{\psi}\overline{\psi}}-\ket{\overline{\psi}\psi\psi}-\ket{\psi\overline{\psi}\psi}-\ket{\psi\psi\overline{\psi}})$. This is a  GHZ state in the diagonal-antidiagonal basis ($\frac{1}{\sqrt{2}}(\ket{\psi}\pm\ket{\overline{\psi}})$).

In a similar fashion, Alice can prepare non-maximally entangled state to two of her partners. She projects her two qubits on states 
$\cos\alpha\ket{\psi} \pm \sin\alpha\ket{\overline{\psi}}$ and $\ket{\Psi_{4}^{-}}$ 
collapses onto $(\cos^2\alpha\ket{\overline{\psi}\overline{\psi}}-\sin^2\alpha\ket{\psi\psi})
/\sqrt{\cos^4\alpha +\sin^4\alpha}$

In Table. \ref{t1}  we give the probabilities (in the ideal cases) of the remote preparations of specific states.
\begin{table}
\caption{Probabilities of RSP for emissions of $\ket{\Psi_k^{-}}$, for $k = 2, 4, 6$.}
\begin{ruledtabular}
\begin{tabular}{lccc}
shared state            &   \# qubits & prepared state &    probability \\ \hline \\[-2ex]
$\ket{\Psi_2^-}$    &   1               & $\ket{\psi}$      &   1/2 \\
$\ket{\Psi_4^-}$    &   2                   &   $\ket{\psi\psi}$        & 1/3   \\
$\ket{\Psi_4^-}$    &   2                   &   $\ket{\Psi^+_2}$      &   1/3 \\
$\ket{\Psi_6^-}$    &   3                   &   $\ket{\psi\psi\psi}$        &   1/4 \\
$\ket{\Psi_6^-}$    &   3                   &   $\ket{W}$/$\ket{\overline W}$   &   1/4 \\
$\ket{\Psi_6^-}$    &   3                   &   $\ket{GHZ}$   &   1/4 \\
\end{tabular}
\end{ruledtabular}
\label{t1}
\end{table}
The preparation probabilities in Table~\ref{t1} can be doubled, if the parties specify in advance that they want to remotely prepare  qubit states on a specific great circle of the Bloch sphere. Then, if the remote qubits are $\bar\psi$, the receivers can rotate their qubits to $\psi$ by applying $\sigma_z$ operations.

Note, that due to a permutation symmetry between Bob, Charlie, and David, the state of their qubits is in a symmetric subspace of the common Hilbert space. For this reason Alice cannot remotely prepare a maximally mixed state for each Partner, as she is unable to remove the correlations arising from the symmetry. Yet, she is able to prepare some mixtures, either by entangling her qubits with an ancilla, or by `tracing out' her qubits (that is ignoring one of the actual results at one of her stations). For instance, if she traces out one of her particles and registers that the other ones are in $\ket{\psi}$, the other parties get an even mixture of $\ket{\overline{\psi}\overline{\psi}\overline{\psi}}\bra{\overline{\psi}\overline{\psi}\overline{\psi}}$ and $\ket{\overline W_3}\bra{\overline W_3}$. If she traces out one more qubit, the mixture shared by the other three observers is of $\ket{\overline{\psi}\overline{\psi}\overline{\psi}}\bra{\overline{\psi}\overline{\psi}\overline{\psi}}$, $\ket{\overline W_3}\bra{\overline W_3}$, and $\ket{W_3}\bra{W_3}$, with respective weights $\frac{1}{4},\frac{1}2,\frac{1}4$. Finally, if Alice simply sends the success signal without any measurement, her partners are left with the balanced mixture of $\ket{\overline{\psi}\overline{\psi}\overline{\psi}}\bra{\overline{\psi}\overline{\psi}\overline{\psi}}$, $\ket{\overline W_3}\bra{\overline W_3}$, $\ket{W_3}\bra{W_3}$, $\ket{\psi\psi\psi}\bra{\psi\psi\psi}$ (which is a separable state). All such processes are occur under the proviso that each partner receives a photon.

\begin{figure}
\includegraphics[width=\columnwidth]{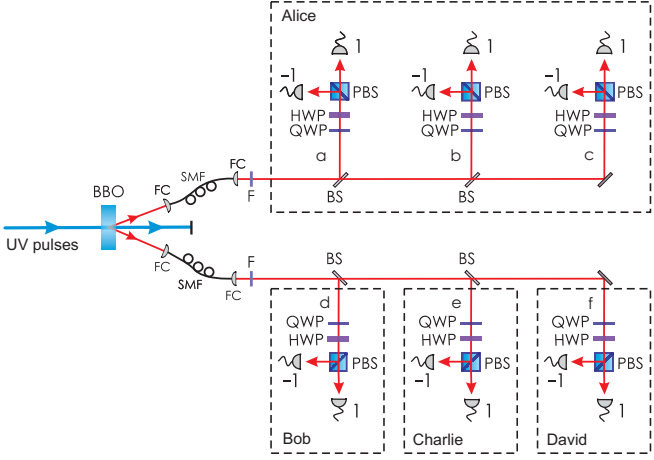}
\caption{\label{setup}
Experimental setup for generating and
analyzing the six-photon polarization-entangled state.  The six
photons are created in third order PDC processes in a 2 mm thick BBO
pumped by UV pulses. The emitted photons are coupled to single mode fibers (SMFs). Narrow band ($\Delta\lambda =3$ nm) interference filters
(Fs) serve to remove spectral distinguishability. The coupled spatial modes are divided into six
exit modes by two pairs 50\%-50\% beam splitters (BSs). Each exit mode can be
analyzed  in arbitrary basis using half- and quarter wave plates
(HWP and QWP) and a polarizing beam splitter (PBS), and two single photon detectors for each mode (the measurement station). The  exit modes $a$, $b$,
and $c$ are controlled Alice, while $d$, $e$, and $f$ by Bob, Charlie, and David, respectively.}
\end{figure}

In our experiment we use a frequency-doubled Ti:Sapphire laser ($80$
MHz repetition rate, $140$ fs pulse length) yielding UV pulses with
a central wavelength at $390$ nm and an average power of $1300$ mW.
The pump beam is focused to a $160$ $\mu$m waist in a $2$ mm thick
BBO ($\beta$-barium borate) crystal. Half wave plates and two $1$ mm
thick BBO crystals are used for compensation of longitudinal and
transversal walk-offs. The emitted photons of non-collinear
type-II PDC are then coupled to single mode fibers (SMFs), defining
the two spatial modes at the crossings of the two frequency
degenerated down-conversion cones. Upon exiting the fibers the
down-conversion light passes narrow band ($\Delta\lambda =3$ nm)
interference filters (Fs) and is split into six spatial modes $(a, b,
c, d, e, f)$ by ordinary $50\%-50\%$ beam splitters (BS), followed
by birefringent optics to compensate phase shifts in the BS's. Due
to the short pulses, narrow band filters, and single mode fibers the
down-converted photons are temporally, spectrally, and spatially
indistinguishable~\cite{ZZW95, WZ, PAN}, see Fig.~\ref{setup}. The
polarization is being kept by passive fiber polarization
controllers. Polarization analysis stations in each exit mode are implemented by a half wave
plate  (HWP), a quarter wave plate (QWP), and a polarizing beam-splitter (PBS) . The outputs of the PBSs are lead to
single photon silicon avalanche photo diodes (APDs) via multi
mode fibers. The APDs' electronic responses, following photo
detections, are being counted by a multi channel coincidence counter
with a $3.3$ ns time window. The coincidence counter registers any
coincidence event between the $12$ APDs as well as single detection
events.

\begin{figure}
 \centering
 \includegraphics[width=\columnwidth]{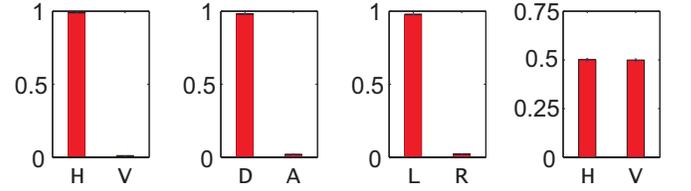}
 \caption{\label{HDL}
The single pair emission case  $\ket{\psi_2^-}$: Renormalized observed detection probabilities for a photon in states $\ket{H}$, $\ket{D}$, $\ket{L}$, and a mixed state at one location for Bob, Charlie, or David (conditional on detection of only one photon by Alice  in specific orthogonal states, see text)}
\end{figure}

\begin{figure}
 \centering
 \includegraphics[width=\columnwidth]{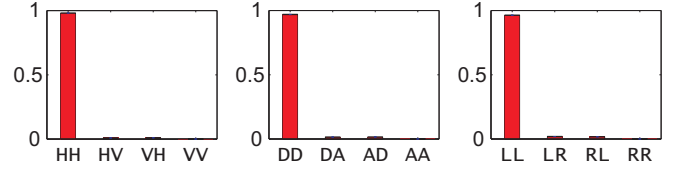}
 \caption{\label{HHDDLL}
The two pair emission case  $\ket{\psi_4^-}$. Renormalized  detection probabilities for two of qubits both in either $\ket{H}$, or $\ket{D}$, or $\ket{L}$ at two locations for Bob and Charlie  (conditional on detection of two photons in specific orthogonal states by Alice, see text).}
\end{figure}

\begin{figure}
 \centering
 \includegraphics[width=0.67\columnwidth]{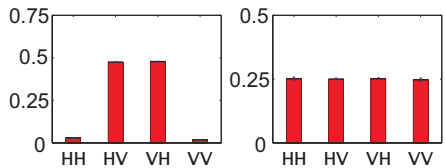}
 \caption{\label{HVVH}
RSP of $\ket{\Psi_2^+}$ and a  two qubit mixed state. Renormalized detection probabilities for two qubit detection events for Bob and Charlie, for the respective case of RSP.}
\end{figure}

\begin{figure}
 \centering
 \includegraphics[width=\columnwidth]{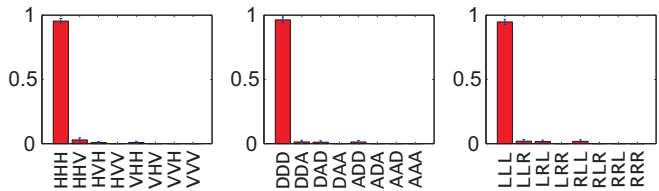}
 \caption{\label{HHHDDDLLL}
RSP of three identical qubit states ($\ket{\psi_6^-}$ emissions).Three photon detection probabilities for the case of $\ket{H}$, $\ket{D}$, and $\ket{L}$ at the  three locations for Bob, Charlie, and David.}
\end{figure}

\begin{figure}
 \centering
 \includegraphics[width=\columnwidth]{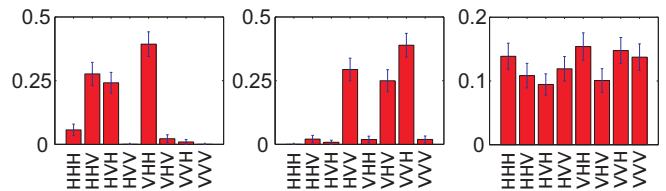}
 \caption{\label{W}RSP of $W$ states. Renormalized detection probabilities for three qubit entangled states $\ket{W}$, $\ket{\overline{W}}$  shared between   Bob, Charlie, and David after a successful RSP. The last graph represents the mixed state defined in the text.}
\end{figure}

The RSP protocol is implemented by projective measurements done by Alice on her qubits. The qubits in exit modes $a$, $b$,
and $c$ are given to Alice, and in each mode one has a polarization measuring station, see Fig. \ref{setup}. The qubit in modes $d$, $e$, and $f$ are given Bob, Charlie, and David,  respectively.  For example, if Alice like to prepare $\ket{H}$ for her three partners, she projects the state of her photons onto $\ket{VVV}$ implies that the remaining three photons are all $\ket{HHH}$. Hence Alice can  in this manner probabilistically prepare qubits in the $\ket{HHH}$ state for her three partners. Due to the probabilistic nature of projective measurements on $\ket{\Psi_{6}^{-}}$, Alice also needs to send classical information indicating the  success to each of her partners, informing them that the intended state has been remotely prepared for them.
In the experiment, we have tested a possibility to prepare horizontally, diagonally, and left circularly polarized photons, as well as the two-qubit maximally entangled states. For two pairs emissions the states which we prepared were $|HH\rangle$, $\ket{DD}$, $\ket{LL}$, $\ket{\psi^+_2}$, as well as $\frac{1}4(\ket{HH}\bra{HH}+\ket{VV}\bra{VV})+\frac{1}{2}\ket{\psi_2^+}\bra{\psi_2^+}$. Finally, for three pair emissions we realized  preparations of $\ket{HHH}$, $\ket{DDD}$, $\ket{LLL}$ and the mixture $\frac{1}{8}(\ket{HHH}\bra{HHH}+\ket{VVV}\bra{VVV})+\frac{3}{8}(\ket{W}\bra{W}+\ket{\bar{W}}\bra{\bar{W}})$.
In figure \ref{HDL}, we show experimental results of three-location RSP of horizontally $H$, diagonally $D$, and  left circularly $L$ polarized photons. The one qubit fidelities are $F_H = 0.98 \pm 0.02$, $F_D = 0.97 \pm 0.04$, and $F_L = 0.97 \pm 0.05$ respectively.
In figures \ref{HHDDLL} and \ref{HVVH} we show experimental results of two-location RSP of horizontally $HH$, diagonally $DD$,   left circularly $LL$ polarized photons, and the two-qubit entangled state $\psi_2^+$ with  fidelities are $F_{HH} = 0.97 \pm 0.04$, $F_{DD} = 0.97 \pm 0.04$, $F_{LL} = 0.97 \pm 0.04$, and $F_{\psi_2^+} = 0.96 \pm 0.03$ respectively.
RSP of  the three-qubit entangled $W$ or $\overline W$ states has been demonstrated by projections at Alice stations to $\ket{HVV}$, $\ket{VHV}$ or $\ket{VVH}$. Similarly,  registrations of $\ket{HHV}$, $\ket{HVH}$ or $\ket{VHH}$ were used to prepare $\overline W$. RSP of three copies of one qubit is obtained by projection of Alive qubits to $\ket{VVV}$. The results are given in figures~\ref{HHHDDDLLL} and \ref{W}.
The three qubit states fidelities are $F_{HHH} = 0.97 \pm 0.07$, $F_{DDD} = 0.97 \pm 0.07$, $F_{LLL} = 0.96 \pm 0.07$, $F_W = 0.90 \pm 0.09$, and $F_{\overline{W}} = 0.91 \pm 0.09$.

The figures clearly show,that we have demonstrated a method to remotely prepare several types of  states of one, two, or three qubits  (product, $\ket{\psi^+}$, $W$, and GHZ). The states are produced by projective measurements on one half of rotationally invariant multipartite states,
 which are readily available in laboratories, via a simple beam-splitting method (which by avoiding interferometric overlaps leads to is stable configuration). Our scheme involves multi-photon interferometry  using a pulsed PDC based source of entangled photons. The experimental data confirm the high precision, with which RSP can work using such experimental  methods.
Interestingly, this scheme works as a, kind of,  symmetrizer of states. If Alice  registers a projection on a product state, her partners obtain a symmetric superposition of the product of states orthogonal to ones, which she observed.

{\bf Acknowledgements} This work was supported by the Swedish Research
Council (VR).  M.\.{Z}. was supported by the Wenner-Gren Foundations
and by the EU program Q-ESSENCE.
M.W. was supported by a CHIST-ERA/NCBiR project QUASAR and Foundation for Polish Science (contract HOMING-PLUS/2011-4/11) and thanks Ryszard Weinar for discussions.\\

\end{document}